\documentstyle[aps,prd]{revtex}
\tighten
\def \D {\mbox{D}}
\def \c {\mbox{curl}\,}

\def \ts {\textstyle}
\def \bq{\begin{equation}}
\def \eq {\end{equation}}
\def \pd {\partial}
\def \bm {\bibitem}
\def \d {\mbox{d}}
\def\rd{\displaystyle{\cdot}}
\def \lb {\label}
\begin{document}

\title{
Stress effects in structure formation
}

\author{
Roy Maartens$^*$\thanks{roy.maartens@port.ac.uk},
Josep Triginer$^{\dag}$\thanks{pep@ulises.uab.es},
and David R. Matravers$^*$\thanks{david.matravers@port.ac.uk}
}

\address{~}

\address{
$^*$\,Relativity and Cosmology Group,
School of Computer Science and Mathematics, University of
Portsmouth, Portsmouth~PO1~2EG, England
}

\address{
$^{\dag}$\, Department of Physics, Autonomous
University of Barcelona, Bellaterra~08193, Spain
}

\address{~}

\date{July 1999}

\maketitle

\begin{abstract}

Residual velocity dispersion in cold dark matter induces stresses
which lead to effects that are absent in the idealized dust model.
A previous Newtonian analysis showed how this approach can provide
a theoretical foundation for the phenomenological adhesion model.
We develop a relativistic kinetic theory generalization which also
incorporates the anisotropic velocity dispersion that will
typically be present. In addition to density perturbations, we
consider the rotational and shape distortion properties of
clustering. These quantities together characterize the linear
development of density inhomogeneity, and we find exact solutions
for their evolution. As expected, the corrections are small and
arise only in the decaying modes, but their effect is interesting.
One of the modes for density perturbations decays less rapidly than
the standard decaying mode. The new rotational mode generates
precession of the axis of rotation. The new shape modes produce
additional distortion that remains frozen in during the subsequent
(linear) evolution, despite the rapid decay of the terms that
caused it.

\end{abstract}


\section{Introduction}

The Cold Dark Matter (CDM) model has had considerable success,
based on using adiabatic perturbations of pressure-free dust on a
Friedman-Robertson-Walker (FRW) background to study the growth of
structure in the matter distribution. The model is simple and the
solutions are easy to interpret (see, e.g., \cite{pd}). The
idealized dust assumption, i.e., exactly zero velocity dispersion,
breaks down when density fluctuations begin to go nonlinear;
caustics and infinite density layers form through shell crossing,
precisely because velocity dispersion is forced to vanish.
Theoretical modifications to complement the extensive numerical
simulations and deal with the multi-stream flow problem are few in
number. One of the most successful is the adhesion model \cite{gu}.
This model has relied on a phenomenological justification rather
than a theoretical derivation. Recently Buchert and Dom\'{\i}nguez
\cite{bd} developed theoretical models which contain the adhesion
one as a special case. Furthermore, their models do not have the
problem of the possible non-conservation of momentum which occurs
in the adhesion model.

In a Newtonian framework, with comoving coordinates on an expanding
background, they use the Poisson-Vlasov equations to obtain
consistent models of a self-gravitating collisionless gas. The
models are designed to allow for a small amount of velocity
dispersion in the gas. The outcome of their approximation scheme is
a system of equations that includes an effective viscosity term
which is more general than the adhesion term, but which can be
specialized to it. They point out that the inclusion of the
velocity dispersion allows access to smaller spatial scales than
previous models permit and could be used to connect studies of
large scale structures with those of smaller ones. This aspect
remains to be investigated.

A well known problem with the Vlasov hierarchy of moment equations
is that it is infinite \cite{bt}, and some additional information
has to be provided. Without collisions, there is in general no
mechanism for eliminating the quadrupole and higher moments. The
simplest approach is to truncate above the dipole and use a dust
model, but this has no velocity dispersion. A physically reasonable
model is obtained in \cite{bd} by assuming small velocity
dispersion, leading to truncation above the quadrupole. This closes
the hierarchy and allows limited velocity dispersion.

In this paper, we develop a relativistic generalization of their
approach, based on the Einstein-Liouville equations. One limitation
of their model is that they assume the velocity dispersion is
isotropic, so that the stress (pressure) is purely isotropic. Our
generalization eliminates the isotropy assumption, allowing for
anisotropic stress.\footnote{
Recently Hu and Eisenstein \cite{he} have investigated anisotropic
stress effects in general, by postulating phenomenological
parametrizations of stress evolution. Our model is of more limited
applicability, but is self-consistent, since the stress evolution
is governed by the kinetic-theory Liouville equation.
}
We are able to find self-consistent (i.e., based on kinetic theory,
rather than ad hoc phenomenology) evolution equations for the
isotropic and anisotropic stresses. Furthermore, we use a covariant
gauge-invariant approach \cite{be,mt} to describe not only the
magnitude of density inhomogeneities, i.e., the density
perturbations, but also their rotational and shape distortion
properties. This leads to a unified system of equations governing
the linear evolution of density inhomogeneity in a physically
realistic model of dark matter. We find the exact solutions of
these equations for a flat background (i.e., $\Omega_{\rm cdm}=1$).
These solutions are relevant for the study of dark matter halo
formation, neglecting baryons and assuming zero cosmological
constant.

Free-streaming effects tend to smooth density fluctuations, while
the energy density supported by stresses can enhance them. The
exact solution shows that the growing mode of density perturbations
is unchanged, while there are two extra decaying modes, one
decaying less rapidly than the standard dust mode, and one more
rapidly. Velocity dispersion will have a purely dissipative effect
on angular momentum, and this is confirmed by the exact solution of
the rotational equation, which shows an extra decaying mode that
decays more rapidly than the standard mode. However, this new mode
has the interesting effect of changing the direction of the axis of
rotation.

The main impact of velocity dispersion is on the evolution of
shape-distortion in in the density distribution. The stresses,
though small and decaying, have a significant effect, producing
`active' distortion in addition to the inertial distortion that
arises in dust models purely from the shear anisotropy. Despite
being sourced by decaying terms, these distortions remain frozen in
during the subsequent evolution (until the nonlinear regime).

In section II we develop the self-consistent kinetic theory
analysis of stress in CDM. Section III presents the covariant
evolution equations for density perturbations, rotation and shape
distortion, and gives the exact solutions of these equations.
Finally, concluding remarks are made in section IV. We follow the
notation of \cite{mt,em}. The signature is $(-+++)$, units are such
that $8\pi G=1=c$ and $k_{\rm B}=1$, spacetime indices are
$a,b,\cdots$, and (square) round brackets enclosing indices denote
(anti-) symmetrization. The spacetime metric is $g_{ab}$, and the
spacetime alternating tensor is
$\eta_{abcd}=-\sqrt{-g}\,\delta_{[a}{}^0
\delta_b{}^1\delta_c{}^2\delta_{d]}{}^3$.

\section{Kinetic model of CDM stresses}

We use the covariant Lagrangian approach to relativistic kinetic
theory \cite{egs,e,em,mw}, in which all the variables are
physically measurable and which allows for a clear Newtonian
interpretation. Given a 4-velocity field $u^a$, we decompose the
4-momentum $p^a$ of a particle of mass $m$ as
\bq
p^a=Eu^a+\lambda^a\,,
\lb{1}
\eq
where $E$ is the particle energy relative to comoving observers,
and
\[
\lambda^a=\lambda e^a=m\gamma(v)v^a
\]
is the particle 3-momentum, with $e_ae^a=1$, $e^au_ a=0$, and
$\lambda=mv(1-v_av^a)^{-1/2}=(E^2-m^2)^{1/2}$. The covariant volume
element in momentum space is
\[
\frac{\d^3\lambda}{E}=\frac{\lambda^2\d\lambda\d\Omega}{E}
= \lambda \d E \d \Omega\,,
\]
where $\d\Omega$ is the solid angle spanned by two independent $\d
e^a$. The distribution function $f(x,E,e^a)$ can be expanded in
tensor multipoles $F_{a_1\cdots a_\ell}(x,E)$, i.e.,
\bq
f=F+F_ae^a+F_{ab}e^ae^b+F_{abc}e^ae^be^c+\cdots\,.
\lb{4a}
\eq
This is the covariant generalization of the spherical harmonic
expansion $f=\sum f_{\ell m}Y_{\ell m}$. The covariant multipoles
are irreducible, i.e., $F_{a\cdots b}=F_{\langle a\cdots
b\rangle}$, where the angled brackets denote the spatially
projected symmetric tracefree (PSTF) part. They are given by
\bq
F_{a_1\cdots a_\ell}=\frac{(2\ell+1)!}{4\pi (\ell!)^22^\ell}
\int fe_{\langle a_1}e_{a_2}\cdots e_{a_\ell\rangle}\d\Omega \,.
\lb{5}
\eq

The energy-momentum tensor is
\begin{eqnarray}
T_{ab} &=& \int fp_ap_b{\d^3\lambda\over E} \nonumber\\
{}&=& \rho u_au_b+ph_{ab}+2q_{(a}u_{b)}+\pi_{ab}\,,
\lb{5a}
\end{eqnarray}
where $h_{ab}=g_{ab}+u_au_b$ is the spatial projector. The energy
density, pressure, energy flux (momentum density), and anisotropic
stress are given by
\begin{eqnarray}
\rho &=& 4\pi\int_m^{\infty}E^2\lambda F\d E\,,\label{m1}\\
p & =& \frac{4\pi}{3}\int_m^{\infty}\lambda^3F\d E \,,\label{m2}\\
q_a  &=& \frac{4\pi}{3}\int_m^{\infty}E\lambda^2F_a\d E \,,\label{m3}\\
\pi_{ab} &=&\frac{8\pi}{15}\int_m^{\infty}\lambda^3F_{ab}\d E
\,.\label{m4}
\end{eqnarray}
Higher-level dynamical anisotropy than $\pi_{ab}$ is defined via
the $\ell\geq3$ multipoles. For example, the octopole anisotropy is
\begin{equation}
\zeta_{abc}={8\pi\over35}\int_m^\infty E\lambda^2F_{abc}\d E\,.
\label{m5}
\end{equation}
The number density is given by \cite{e}
\[
n=4\pi\int_m^\infty E\lambda F\d E\,,
\]
and we can derive a useful relation between the monopole dynamical
terms (compare the similar relation found in \cite{mw}):
\begin{equation}
mn+{\ts{3\over2}}p=\rho-{\ts{1\over2}}
{\cal M}~\mbox{ where }~{\cal
M}=4\pi\int_m^\infty\left(1-{m\over E}\right)^2E^2\lambda F\d
E\,. \label{m6}
\end{equation}
In the massless limit $m\rightarrow 0$, we have ${\cal
M}\rightarrow\rho$, and Eq. (\ref{m6}) reduces to
$p={1\over3}\rho$. Equation (\ref{m6}) is a generalized `equation
of state', which adopts a simple form in the massless limit and the
limit of low velocity dispersion (see below), but which is more
complicated in intermediate regimes.

The energy-momentum conservation equations $\nabla^bT_{ab}=0$ follow
from the Liouville (collisionless Boltzmann) equation, and are
\cite{e}
\begin{eqnarray}
\dot\rho+(\rho+p)\Theta+\D^aq_a &=&
-2A^aq_a-\sigma^{ab}\pi_{ab}\,,
\label{c1}\\
\dot{q}_{\langle a\rangle}+{\ts{4\over3}}\Theta
q_a+(\rho+p)A_a+\D_ap+\D^b\pi_{ab}
&=&-\sigma_{ab}q^b+\varepsilon_{abc}\omega^bq^c-A^b\pi_{ab}\,.
\label{c2}
\end{eqnarray}
Here $\D_a$ is the spatially projected covariant derivative, i.e.,
\[
\D_aS_{b\cdots c}=h_a{}^dh_b{}^e\cdots h_c{}^f\nabla_dS_{e\cdots f}\,,
\]
an overdot is the covariant time derivative, i.e.,
$\dot{S}_{a\cdots b}= u^c\nabla_cS_{a\cdots b}$, and
$\varepsilon_{abc}=\eta_{abcd}u^d$ is the spatial alternating
tensor. The expansion, acceleration, vorticity and shear of the
4-velocity $u^a$ are given by
\[
\Theta=\D^au_a\,,~A_a=\dot{u}_a\,,~\omega_a=-{\ts{1\over2}}\c u_a\,,~
\sigma_{ab}=\D_{\langle a}
u_{b\rangle}\,.
\]
The covariant spatial curl of vectors and rank-2 tensors is defined
by \cite{m}
\[
\c V_a=\varepsilon_{abc}\D^bV^c\,,~\c S_{ab}=
\varepsilon_{cd(a}\D^cS^d{}_{b)}\,.
\]

We are free to choose the 4-velocity $u^a$ so that $q_a=0$, i.e.,
so that in the comoving frame, no energy flux is observed
\cite{e,is}. In general, there will be a non-vanishing particle
drift in this frame. To maintain vanishing energy flux, the
momentum conservation equation (\ref{c2}) shows that
\begin{equation}
(\rho+p)A_a+\D_ap+\D^b\pi_{ab}
=-A^b\pi_{ab}\,.
\label{c2'}
\end{equation}
Thus the evolution equation (\ref{c2}) for $q_a$ becomes a
constraint equation (\ref{c2'}) for the acceleration. Note also
that in the energy frame, the dipole $F_a$ will satisfy $\int
E\lambda^2F_a\d E=0$, from Eq. (\ref{m3}). From now on, we assume
that the energy frame is chosen, i.e. $q_a=0$.

In a universe that is close to an FRW model, i.e. with small
inhomogeneity and anisotropy, we have that \cite{be,cmb}
\[
{A_a\over\Theta}\,,{\omega_a\over\Theta}\,,{\sigma_{ab}\over\Theta}\,,
{\pi_{ab}\over\rho}\,,{a\D_a p\over\rho}\,,{a\D_a \rho\over\rho}\,,
{a\D_a \Theta\over\Theta}={\cal O}(\epsilon)\,,
\]
where $\epsilon$ is a dimensionless smallness parameter, and $a$
is the cosmic scale factor in the FRW background.
(In the background, $\Theta=3H$, where $H$ is the Hubble rate.)
The higher-level dynamical anisotropy tensors are also
${\cal O}(\epsilon)$:
\[
{\zeta_{abc}\over\rho}\,, \cdots ={\cal O}(\epsilon)\,.
\]
To linear order in such a universe, the conservation equations
(\ref{c1}) and (\ref{c2'}) reduce to
\begin{eqnarray}
\dot\rho+(\rho+p)\Theta &=&0\,,
\label{c1'}\\
(\rho+p)A_a+\D_ap+\D^b\pi_{ab}
&=&0\,.
\label{c2''}
\end{eqnarray}
From now on, we will consider a universe that is close to FRW,
i.e., we drop all ${\cal O}(\epsilon^2)$ terms. The Liouville
equation may be decomposed into multipole evolution equations that
are PSTF \cite{em}. The monopole, dipole and quadrupole evolution
equations are
\begin{eqnarray}
E\dot{F}+{\ts{1\over3}}\lambda\D^aF_a-
{\ts{1\over3}}\lambda^2\Theta\frac{\pd
F}{\pd E} &=&0\,,
\lb{10}\\
E\dot{F}_a+{\ts{2\over5}}\lambda\D^bF_{ab}-
{\ts{1\over3}}{\lambda^2}\Theta
\frac{\pd F_a}{\pd E}+\lambda\D_aF-\lambda E\frac{\pd F}{\pd
E}A_a &=&0 \,,
\lb{11}\\
E\dot{F}_{ab}+{\ts{3\over7}}\lambda\D^cF_{abc}-
{\ts{1\over3}}{\lambda^2}\Theta
\frac{\pd F_{ab}}{\pd E}+\lambda\D_{\langle a}F_{b\rangle }-
\lambda^2\frac{\pd F}{\pd E}
\sigma_{ab} &=&0 \,.
\lb{12}
\end{eqnarray}
(Note that the vorticity does not enter the Liouville multipoles at
the linear level.) Multiplying Eq. (\ref{10}) by $E\lambda$ and
integrating over all energies, and using the energy frame condition
$\int E\lambda^2F_a\d E=0$, we derive the energy conservation
equation (\ref{c1'}). Similarly, multiplying Eq. (\ref{11}) by
$\lambda^2$ and integrating, we arrive at the momentum conservation
equation (\ref{c2''}). When integrating by parts to obtain some of
these terms, we use the assumption that as $E\rightarrow\infty$,
$F_{a_1\cdots a_\ell}$ ($\ell\geq0$) tends to zero more rapidly
than $E^n$ for any $n<0$.

We can derive a new evolution equation for the pressure after multiplying the
monopole equation (\ref{10}) by $\lambda^3/E$:
\begin{equation}
\dot{p}+{\ts{5\over3}}\Theta p={\ts{1\over3}}\Theta{\cal
P}-{\ts{1\over3}}\D^a{\cal Q}_a\,, \label{c3}
\end{equation}
where
\begin{eqnarray*}
{\cal P} &=& {4\pi\over3}\int_m^\infty\left(1-{m^2\over
E^2}\right)\lambda^3F\d E \,,\\
{\cal Q}_a &=& {4\pi\over3}\int_m^\infty\left(1-{m^2\over
E^2}\right)E\lambda^2F_a\d E \,.
\end{eqnarray*}
In the massless limit, Eq. (\ref{c3}) reduces to the energy
conservation equation. But in general, Eq. (\ref{c3}) is a new and
nontrivial evolution equation arising from the Liouville equation.

A new evolution equation for the anisotropic stress $\pi_{ab}$ may
also be found after multiplying the quadrupole equation (\ref{12})
by $\lambda^3/E$:
\bq
\dot{\pi}_{ab}+{\ts{5\over3}}\Theta\pi_{ab}+2p\sigma_{ab}
=-{\ts{2\over5}}{\cal P}\sigma_{ab}
-{\ts{2\over5}}\D_{\langle a}{\cal Q}_{b\rangle}+{\ts{1\over3}}
\Theta {\cal R}_{ab}-\D^c{\cal S}_{abc}\,,
\lb{12''}\eq
where
\begin{eqnarray*}
{\cal R}_{ab} &=& {8\pi\over15}\int_m^\infty\left(1-{m^2\over
E^2}\right)\lambda^3F_{ab}\d E \,,\\
{\cal S}_{abc} &=& {8\pi\over35}\int_m^\infty\left(1-{m^2\over
E^2}\right)E\lambda^2F_{abc}\d E \,.
\end{eqnarray*}
In the massless limit, we have ${\cal Q}_a\rightarrow q_a \,(=0)$,
${\cal R}_{ab}\rightarrow\pi_{ab}$, ${\cal
S}_{abc}\rightarrow\zeta_{abc}$, and Eq. (\ref{12''}) reduces to
the evolution equation for free-streaming radiation that was found
in \cite{cmb}.

The Liouville multipole equations (\ref{10})--(\ref{12}) are the
beginning of an infinite hierarchy. (See \cite{cmb,cmb2} for the
corresponding equations in the massless case, which is much
simpler.) The evolution equation (\ref{12''}) for anisotropic
stress contains the spatial divergence of the octopole, and the
octopole evolution equation will contain the divergence of the
hexadecapole, and so on. In general, the evolution equation for the
$\ell$-pole has the spatial divergence of the $(\ell+1)$-pole as an
effective source term, so that power is transmitted across levels
of the hierarchy. Thus the multipoles above the quadrupole affect
dynamical evolution, even though they do not directly enter the
Einstein field equations. The Liouville hierarchy cannot in general
be truncated, without some approximation scheme to close the
truncated system. For a collisional gas, one expects on physical
grounds that interactions tend to thermalize, and the higher
multipoles will tend to be suppressed. For a collisionless and
massless gas, anisotropy in the higher multipoles does not in
general decay through free-streaming in an expanding universe,
since the velocity of particles is not affected by redshifting. On
the other hand, redshifting the momentum of massive particles
reduces the peculiar velocity $v$.

Up to this point, our results apply to any collisionless gas in a
nearly FRW universe. Now we need to specialize to the case of CDM,
for which the velocity dispersion is small. This allows us to
develop a consistent approximation scheme for truncating the
Liouville hierarchy, following an approach similar to that of
\cite{bd}. Small velocity dispersion means that there is a small
effective maximum velocity $v_*$, above which the distribution is
effectively vanishing. More precisely,
\[
v_*^2={\cal O}(\epsilon)~\mbox{ and }~{1\over\rho}\int_{E_*}^\infty
E^{2-n}\lambda^{n+1}F_{a_1\cdots a_\ell}\d E={\cal
O}(\epsilon^2)~\mbox{ for }~\ell\geq0\,,n=0,1,2\,.
\]
We assume that the derivatives of the distribution multipoles are
similarly restricted. With the small velocity dispersion
approximation, we can show that many of the terms in the equations
above are second-order. For example,
\begin{eqnarray*}
p &=&{4\pi\over3}\int_m^{\infty}{\lambda^2\over E^2}\,E^2\lambda F\d
E \\
&=&{4\pi\over3}\int_m^{\infty}\left[v^2+{\cal
O}(v^4)\right]E^2\lambda F\d E\\
&\leq &{\ts{1\over3}}v_*^2\left(4\pi\int_m^\infty E^2\lambda F\d
E\right)+\rho{\cal O}(\epsilon^2)\,,
\end{eqnarray*}
so that
\[
{p\over\rho}\leq {\ts{1\over3}}v_*^2+{\cal O}(\epsilon^2)\,.
\]
Similarly, we find that
\begin{eqnarray*}
{{\cal M}\over\rho} &\leq & v_*^4+{\cal O}(\epsilon^2)\,,\\
{{\cal P}\over\rho} &\leq & {\ts{1\over3}}v_*^4+{\cal O}(\epsilon^2)\,,\\
{|{\cal Q}_a|\over\rho} &\leq & v_*^2{|q_a|\over\rho}+{\cal O}(\epsilon^2)\,,\\
{|{\cal R}_{ab}|\over\rho} &\leq & v_*^2{|\pi_{ab}|\over\rho}
+{\cal O}(\epsilon^2)\,,\\
{|{\cal S}_{abc}|\over\rho} &\leq & v_*^2{|\zeta_{abc}|\over\rho}
+{\cal O}(\epsilon^2)\,.
\end{eqnarray*}
To linear order, it follows that Eq. (\ref{m6}) produces the
equation of state
\begin{equation}
\rho=mn+{\ts{3\over2}}p\,, \label{m6'}
\end{equation}
which simply expresses that each particle has rest mass $m$ and
kinetic energy ${1\over2}mv^2$ to lowest order. The stress
evolution equations (\ref{c3}) and (\ref{12''}) reduce to
\begin{eqnarray}
\dot{p}+{\ts{5\over3}}\Theta p&=&0\,,\label{19a}\\
\dot{\pi}_{ab}+{\ts{5\over3}}\Theta\pi_{ab} &=&0 \,. \label{19b}
\end{eqnarray}
Since $p/\rho={\cal O}(\epsilon)$, the term $p\sigma_{ab}$ is
second order and falls away from the stress evolution equation
(\ref{19b}). The octopole anisotropy does not contribute to the
stress evolution at linear order, so that the multipole hierarchy
can be truncated after the quadrupole. The higher-multipole
evolution equations are decoupled from the Einstein-Liouville
system at linear order. Equations (\ref{c1'}), (\ref{19a}) and
(\ref{19b}) form a closed system of evolution equations for the
dynamical quantities $\rho$, $p$ and $\pi_{ab}$.

Our approximation scheme extends that of \cite{bd} from a Newtonian
to a relativistic treatment, but it also generalizes the
description of the matter. In \cite{bd}, it is assumed that
$\pi_{ab}=0$, implying the very restrictive condition of {\em
isotropic velocity dispersion}. We do not make this assumption; on
the contrary, the anisotropic stress $\pi_{ab}$ plays a crucial
role in our analysis.

There are some formal similarities here to the Grad 14-moment
method as applied in the hydrodynamic near-equilibrium regime. In
that context, the Boltzmann hierarchy is also truncated beyond the
quadrupole, and anisotropic stress obeys the Israel-Stewart
transport equation \cite{is}
\[
\tau\dot{\pi}_{ab}+\pi_{ab}=-2\eta\sigma_{ab}\,,
\]
where $\tau$ is a relaxation timescale, and $\eta$ is the shear
viscosity. This transport equation has a similar form to our
equation (\ref{19b}). However, the Israel-Stewart transport
equation, and the relativistic Grad method which it is based on,
apply to a {\em collision-dominated} gas, whereas we are dealing
with a collision-free gas.

Note that since $p/\rho={\cal O}(\epsilon)$, the momentum
constraint equation (\ref{c2''}) reduces to
\begin{equation}
\rho A_a+\D_ap+\D^b\pi_{ab}=0\,.
\label{11''}
\end{equation}
In the background ($\epsilon\rightarrow 0$), we have $p\rightarrow
0$. This means that the background distribution function reduces to
a delta-function, since there is no velocity dispersion, and we
have the kinetic theory form of the dust model \cite{e}. In the
inhomogeneous perturbed universe, the monopole $F$ of the
distribution function is not a delta-function, since there is
velocity dispersion. Thus perturbation of the background not only
produces nonzero dipole and higher multipoles, but also changes the
monopole.

\section{Covariant analysis of density inhomogeneity}

In this section we provide the basic equations governing the
evolution of density inhomogeneity in cold dark matter when
isotropic and anisotropic stresses are incorporated. The full set
of covariant and gauge-invariant perturbation equations for a
general energy-momentum tensor is derived and discussed in
\cite{be,mt}. The formalism is based on constructing covariant
quantities which vanish in the background, thus ensuring that they
are gauge-invariant. Density inhomogeneity is described by the
comoving fractional density gradient (which fulfils the above
requirements):
\bq
\delta_a=\frac{a\D_a\rho}{\rho}\,,
\lb{21a}
\eq
where $a$ is the background scale factor.

This quantity carries information about the magnitude, rotational
and shape-distortion properties of inhomogeneity, obtained by
irreducibly splitting its comoving gradient \cite{be}:
\bq
a\D_b\delta_a=({\ts{1\over3}}\delta) h_{ab}
+\varepsilon_{abc}W^c +\xi_{ab}\,.
\lb{21b}
\eq
Here
\[
\delta\equiv a\D^a\delta_a={(a\D)^2\rho\over\rho}
\]
corresponds to the gauge-invariant density perturbation scalar
$\epsilon_m$ in the metric-based formalism \cite{ba}. The quantity
\[
W_a=-{\ts{1\over 2}}a\,\c \delta_a
\]
describes the rotational properties of inhomogeneous clustering,
and it is proportional to the vorticity $\omega_a$. Finally,
\[
\xi_{ab}=a\D_{\langle a}\delta_{b\rangle }
\]
describes the volume-true distortion of inhomogeneous clustering.

These quantities completely and covariantly describe infinitesimal
inhomogeneities in the density. They obey evolution equations in
which the stresses are source terms. Since the pressure $p$ is
${\cal O}(\epsilon)\rho$, we may neglect it in the background. We
assume a flat (i.e., Einstein-de Sitter) background, neglecting the
baryonic component. Thus we are investigating density inhomogeneity
in CDM in the linear regime, with potential applications to dark
matter halo formation. The background field equations give
\bq
\rho=3H^2\,,~H={2\over3t}\,,~a=a_0\left({t\over t_0}\right)^{2/3}\,.
\lb{20'}\eq

The evolution equations (\ref{19a}) and (\ref{19b}) for CDM
stresses in a nearly FRW universe can be integrated to give
\begin{eqnarray}
\pi_{ab} &=& \pi^{(0)}_{ab}\left(\frac{a_0}{a}\right)^5\,,\lb{25}\\
p &=& p_0\left(\frac{a_0}{a}\right)^5 \,,
\lb{20}
\end{eqnarray}
where
\[
\dot{\pi}^{(0)}_{ab}=0=\dot{p}_{0}\,.
\]
Using the linearized identity
\[
\left(a\D_aS_{b\cdots c}\right)^{\rd}=a\D_a\dot{S}_{b\cdots c}\,,
\]
which holds for any tensor $S_{a\cdots b}$ that vanishes in the
background, it follows that
\bq
\left(a^n\D_{a_1}\cdots\D_{a_n}\pi^{(0)}_{ab}\right)^{\rd}=0=
\left(a^n\D_{a_1}\cdots\D_{a_n}p_0\right)^{\rd} \,,
\lb{25'}\eq
for any positive integer $n$.

\subsection{Density perturbations}

We consider first the effect of stresses on density perturbations.
The evolution equation for $\delta$, as given by Eq. (28) of
\cite{mt}, reduces to
\bq
\ddot{\delta}+2H\dot{\delta}-{\ts{3\over2}}H^2\delta=
\frac{a^2}{\rho}\D^2(\D^2p)+3H\dot{S}-3H^2S+\D^2S\,,
\lb{22}
\eq
where the anisotropic stress term is
\bq
S\equiv\frac{a^2\D^a\D^b\pi_{ab}}{\rho}\,,
\lb{23}
\eq
and $H$ and $\rho$ are given by Eq. (\ref{20'}). Note that the
isotropic stress $p$ occurs only via the gradient term $\D_a p$.
Using Eqs. (\ref{25})--(\ref{20'}), we find that
\[
S={\ts{3\over4}}t_0^2S_0\left({a_0\over a}\right)^2\,,~
{a^2\over\rho}\D^4p={\ts{3\over4}}\left({t_0\over a_0}\right)^2
P_0\left({a_0\over a}\right)^4
\]
where $\dot{S}_0=0=\dot{P}_0$ and
\bq
S_0\equiv a^2\D^a\D^b\pi^{(0)}_{ab}\,,~
P_0\equiv a^4\D^4p_0 \,.
\lb{26}\eq
Thus the evolution equation (\ref{22}) becomes
\[
\ddot{\delta}+\left({4\over 3t}\right)\dot\delta-\left(
{2\over 3t^2}\right)\delta=
\frac{3}{4}\left({t_0\over a_0}\right)^2(P_0+R_0)
\left({t_0\over t}\right)^{8/3}-
3S_0\left(\frac{t_0}{t}\right)^{10/3}\,,
\]
where
\bq
R_0\equiv a^2\D^2S_0\,,
\lb{27}\eq
so that $\dot{R}_0=0$ by Eq. (\ref{25'}). The solution is
\begin{eqnarray}
\delta &=& C^{(+)}\left({t\over t_0}\right)^{2/3}+C^{(-)}\left({t
\over t_0}\right)^{-1} \nonumber\\
{}&&
-\left[3t_0^2\left({3t_0\over 4a_0}\right)^2
(P_0+R_0)\right]\left({t\over t_0}\right)^{-2/3}
-\left[{\ts{9\over2}}t_0^2S_0\right]\left({t
\over t_0}\right)^{-4/3}\,,
\lb{30}\end{eqnarray}
where $\dot{C}^{(\pm)}=0$.

The standard dust solution is given by the growing $C^{(+)}$ and
decaying $C^{(-)}$ terms. The effects of stress (sourced in
velocity dispersion) are encoded in the following two decaying
terms. Note that one of the two new decaying modes decays less
rapidly than the standard decaying mode, and the other decays more
rapidly. The second, more rapidly decaying, term, is a purely
anisotropic stress term, whereas the first term has isotropic
($P_0$) and anisotropic ($R_0$) stress contributions. When velocity
dispersion is forced to vanish exactly in the dust model, it is
possible to remove the decaying mode by choosing $C^{(-)}=0$. When
velocity dispersion is incorporated, it is no longer possible to
remove decaying modes by choice of initial conditions. This is
related to the fact that the perturbations are no longer adiabatic,
given that the stresses are neglected in the background. The new
decaying terms depend on the initial spatial distribution of
stresses, as described by the quantities $P_0$, $R_0$ and $S_0$,
defined in Eqs. (\ref{26}) and (\ref{27}). By the momentum
conservation equation (\ref{11''}), we can replace $P_0+R_0$ by a
term proportional to the Laplacian of the divergence of the
4-acceleration:
\[
P_0+R_0=-\rho_0a_0\left({t\over
t_0}\right)^{2}(a\D)^2\left(a\D^aA_a\right)\,.
\]

\subsection{Rotational instability}

The evolution equation for the rotational part $W_a$ of density
inhomogeneity is given in \cite{mt}:
\bq
\dot{W}_a+{\ts{3\over2}}H{W_a}
=-\left(\frac{3H}{2\rho}\right)a^2\c\D^b\pi_{ab} \,.
\lb{40}
\eq
Using Eqs. (\ref{25})--(\ref{20'}), this becomes
\bq
\dot{W}_a+\left({1\over t}\right){W_a}=
-{\ts{3\over4}}t_0N_a\left({t_0\over t}\right)^{7/3}\,,
\lb{44}
\eq
where
\bq
N_a\equiv a^2\c\D^b\pi^{(0)}_{ab}\,,
\lb{44'}\eq
so that $\dot{N}_a=0$. Note that we can use the linearized form of
the differential identities in \cite{m} to rewrite this as
\[
N_a=2a^2\D^b\c\pi^{(0)}_{ab}\,.
\]

The solution of Eq. (\ref{44}) is
\bq
W_a=C^{(-)}_a\left({t\over t_0}\right)^{-1}+
\left[{\ts{9\over4}}t_0^{2}N_a\right]\left({t\over t_0}
\right)^{-4/3} \,,
\lb{45}
\eq
where $\dot{C}^{(-)}_a=0$. The standard dust solution is the
$C^{(-)}_a$ term, and the effect of velocity dispersion is to
introduce another decaying mode, which decays more rapidly. The
main effect of this new mode is to break the constancy of direction
of the axis of rotation. It follows from Eq. (\ref{45}) that
\[
(W_a)\big|_{t_0}=C^{(-)}_a+{\ts{9\over4}}t_0^2N_a\,,~
(\dot{W}_a)\big|_{t_0}=-t_0^{-1}\left[C^{(-)}_a+3t_0^2N_a\right]\,.
\]
In the absence of anisotropic stress (or for anisotropic stress
with curl-free divergence), $\dot{W}_a$ remains parallel to $W_a$,
and the direction of the axis of rotation is constant along $u^a$.
When anisotropic stresses are incorporated, $N_a\neq0$ in general,
so that $\dot{W}_a$ is no longer parallel to $W_a$, and the
direction of the axis evolves in time.

\subsection{Shape distortion}

From \cite{mt}, the shape distortion part $\xi_{ab}$ obeys the
evolution equation
\begin{eqnarray}
&&\ddot{\xi}_{ab}+2H\dot{\xi}_{ab}-{\ts{3\over2}}H^2\xi_{ab}
\nonumber\\
{}&&=
\frac{a^2}{\rho}\D_{\langle a}\D_{b\rangle }\D^2p+
\frac{a^2}{\rho}\left[3H\D_{\langle a}\D^c\dot{\pi}_{b\rangle
c}+6H^2\D_{\langle a}\D^c\pi_{b\rangle c}+
\D_{\langle a}\D_{b\rangle }\D^c\D^d\pi_{cd}\right]\,.
\lb{31}
\end{eqnarray}
Using again Eqs. (\ref{25})--(\ref{20'}), we find that
\begin{eqnarray}
&&\ddot{\xi}_{ab}+\left(\frac{4}{3t}\right)
\dot{\xi}_{ab}-\left(\frac{2}{3t^2}\right)\xi_{ab}
\nonumber\\
{}&&=
-\frac{3}{4}\left({t_0\over a_0}\right)^2(P_{ab}+R_{ab})
\left({t_0\over t}\right)^{8/3}-3S_{ab}
\left(\frac{t_0}{t}\right)^{10/3}\,,
\lb{38}
\end{eqnarray}
where we have defined
\bq
P_{ab}\equiv a^4\D_{\langle a}\D_{b\rangle}\D^2p_{0}\,,~
R_{ab}\equiv a^4\D_{\langle a}\D_{b\rangle}\D^c\D^d\pi^{(0)}_{cd}\,,~
S_{ab}\equiv a^2\D_{\langle a}\D^c\pi^{(0)}{}_{b\rangle c}\,,
\lb{38'}\eq
so that
\[
\dot{P}_{ab}=\dot{R}_{ab}=\dot{S}_{ab}=0\,.
\]

Then, as in the scalar case, Eq. (\ref{38}) can be solved to give
\begin{eqnarray}
\xi_{ab} &=&
C^{(+)}_{ab}\left({t\over t_0}\right)^{2/3}
+C^{(-)}_{ab}\left({t\over t_0}\right)^{-1}
\nonumber\\
{}&&
-\left[3t_0^2\left({3t_0\over 4a_0}\right)^2
\left(P_{ab}+R_{ab}\right)\right]\left({t\over t_0}\right)^{-2/3}
-\left[{\ts{9\over2}}t_0^2S_{ab}\right]\left({t\over t_0}
\right)^{-4/3}\,,
\lb{39}
\end{eqnarray}
where $\dot{C}^{(\pm)}_{ab}=0$ . Again we see the occurrence of new
decaying modes arising from stress effects. One of the new terms
decays more slowly than the standard decaying term which arises in
the dust case.

These new terms have the following important implication. We
consider an initially isotropic infinitesimal fluctuation at a
point $\vec{x}_0$, and follow its evolution along $u^a$. The
initial velocity is described via the PSTF and constant tensor
$V_{ab}$, i.e.
\bq
\xi_{ab}(t_0,\vec{x}_0)=0\,,~~
\dot{\xi}_{ab}(t_0,\vec{x}_0)=H_0V_{ab}\,.
\lb{39'}\eq
Let $\tau\equiv t/t_0$, and define the constant PSTF tensors
\[
J_{ab}\equiv {\ts{3\over 80}}t_0^2\left({t_0\over a_0}\right)^2\left[
P_{ab}(\vec{x}_0)+R_{ab}(\vec{x}_0)\right]\,,
~~ K_{ab}\equiv {\ts{1\over10}}t_0^2S_{ab}(\vec{x}_0)\,.
\]
Then Eq. (\ref{39}) gives
\begin{eqnarray}
\xi_{ab}&=&\left[{\ts{2\over5}}\tau^{2/3}
\left(1-\tau^{-5/3}\right)\right]V_{ab} \nonumber\\
&&{}+\left[\tau^{2/3}\left(4-45\tau^{-4/3}+41\tau^{-5/3}\right)\right]J_{ab}
+\left[\tau^{2/3}\left(-4+49\tau^{-5/3}-45\tau^{-2}\right)\right]K_{ab}\,.
\lb{39.}
\end{eqnarray}

Thus for a dust model, in which $J_{ab}=0=K_{ab}$, the evolution of
shape distortion is purely inertial, i.e., it is fixed by the
initial velocity ellipsoid $V_{ab}$, and no further distortion can
develop as the fluctuation evolves (compare \cite{bs}). All the
covariant time derivatives of $\xi_{ab}$ are proportional to
$V_{ab}$:
\bq
\xi_{ab}\propto\dot{\xi}_{ab}\propto\ddot{\xi}_{ab}\propto\cdots\propto
V_{ab}\,.
\lb{39-}\eq

By contrast, when velocity dispersion is incorporated via stress
effects, the same initial conditions in Eq. (\ref{39'}) lead to a
non-trivial evolution of distortion, away from that initially
determined by the velocity ellipsoid $V_{ab}$. The simple relation
in Eq. (\ref{39-}) is broken, and the evolution of distortion is no
longer fixed by the initial velocity ellipsoid. Although the stress
terms that cause the additional `non-inertial' distortion are small
and decaying, once the extra distortion is introduced, there is no
mechanism for removing it, at least during the linear regime. Thus
the distortion is frozen in during the subsequent linear evolution.

The impact of stress on shape distortion is reminiscent of the
impact of stress on shear decay: for radiative anisotropic stress,
Barrow and Maartens \cite{bm} have shown that the decay of shear
due to expansion is slowed down. We expect that the same
qualitative result holds in the case of non-radiative anisotropic
stress, such as considered here.

\section{Conclusion}

Density perturbation theory for the growth of structures in a CDM
framework has been generalized in a covariant form which
self-consistently incorporates small velocity dispersion. The
analysis generalizes the Newtonian approach of Buchert and
Dom\'{\i}nguez \cite{bd} to general relativity; furthermore, it
dispenses with their isotropic dispersion assumption, and considers
the rotational and shape distortion properties of density
inhomogeneity, in addition to the density perturbations. The
evolution equations are integrated exactly for all these parts of
density inhomogeneity in the linear regime.

As a special case ($\pi_{ab}=0$), our results contain the
generalization of the adhesion model, as shown in \cite{bd}. More
generally, our solutions show explicitly how the decaying modes are
modified by stress effects induced via velocity dispersion. These
modifications are small, but they have some important
implications.\\ (1) First, as argued in \cite{bd}, the presence of
velocity dispersion avoids some of the problems that arise in the
dust model, which is pathological in enforcing strictly zero
dispersion. \\ (2) Second, the new decaying modes of density
perturbations reflect non-adiabatic features introduced by the
stresses. One of these modes decays less rapidly than the standard
decaying mode.\\ (3) Third, the new decaying mode in the rotational
part of density inhomogeneity has the effect of breaking the
constancy of the direction of rotation axis.\\ (4) Fourth, the new
decaying modes in the shape distortion mean that additional
`non-inertial' distortion is generated, which is not present in the
dust (purely inertial) model. The additional distortion remains
frozen in during the linear regime, despite the decaying nature of
the source terms, since there is no (linear) mechanism to reverse
it. Although the dominant distortion effects will take place in the
nonlinear regime, this linear effect has some interest, and it may
be worth investigating the statistics of the phenomenon in order to
be able to make more general assertions about the distortion
conditions at the onset of nonlinear structure formation.

The stress effects on rotational and shape-distortion properties of
the density distribution are qualitatively similar to the effects
of a magnetic field \cite{tm}.

\[ \]
{\bf Acknowledgments:}

Thanks to Marco Bruni for incisive comments on fundamental aspects.
RM and JT are partially supported by a European Science Exchange
Programme grant.

\end{document}